\documentclass[prd,superscriptaddress,a4paper,showpacs,showkeys,10pt,nofootinbib]{revtex4}
\usepackage{graphicx}
\usepackage{graphics}
\usepackage{epsfig}
\usepackage{dcolumn}
\usepackage{bm}
\usepackage{multirow}
\usepackage{tabularx}
\usepackage{hyperref}
\usepackage{commath}
\usepackage{epstopdf}
\usepackage[T1]{fontenc}
\usepackage{geometry}
\usepackage{color}

\providecommand{\superchic}{{\sc SuperCHIC}}

\providecommand{\hepmc}{{\sc HepMC}}

\providecommand{\analysisroot}{{\sc ROOT}}
\providecommand{\heppdt}{{\sc HepPDT}}

\def\be{\begin{equation}}
\def\ee{\end{equation}}

\providecommand{\ee}{e$^+$e$^-$}

\def\gappeq{\mathrel{\rlap {\raise.5ex\hbox{$>$}}
{\lower.5ex\hbox{$\sim$}}}}

\def\lappeq{\mathrel{\rlap{\raise.5ex\hbox{$<$}}
{\lower.5ex\hbox{$\sim$}}}}

\def\Toprel#1\over#2{\mathrel{\mathop{#2}\limits^{#1}}}

\def\pom{{I\!\!P}}

\providecommand{\tabularnewline}{\\}

\makeatother

\begin{document}
%
%
\title{Diffractive $\gamma \gamma$ production in $pp$ collisions at the LHC}


\author{V.~P.~Gon\c calves}

\email[]{barros@ufpel.edu.br}

\affiliation{Instituto de F\'{\i}sica e Matem\'atica, Universidade Federal de
Pelotas (UFPel),\\
Caixa Postal 354, CEP 96010-090, Pelotas, RS, Brazil}

\author{D.~E.~Martins}

\email[]{dan.ernani@gmail.com}

\affiliation{Instituto de F\'isica, Universidade Federal do Rio de Janeiro (UFRJ), 
Caixa Postal 68528, CEP 21941-972, Rio de Janeiro, RJ, Brazil}

\author{M.~S.~Rangel}

\email[]{rangel@if.ufrj.br}

\affiliation{Instituto de F\'isica, Universidade Federal do Rio de Janeiro (UFRJ), 
Caixa Postal 68528, CEP 21941-972, Rio de Janeiro, RJ, Brazil}



\begin{abstract}
In this letter we estimate the contribution of the  double diffractive processes for the diphoton production in $pp$ collisions at the Large Hadron Collider (LHC). The acceptance of the central and forward LHC detectors is taken into account and predictions for the invariant mass, rapidity and, transverse momentum distributions are presented. A comparison with the predictions for the  Light -- by -- Light (LbL) scattering and exclusive diphoton production is performed.
We demonstrate that the events associated to double diffractive processes can be separated and its study can be used to constrain the behavior of the diffractive parton distribution functions.
\end{abstract}




\maketitle

\begin{figure}
\begin{tabular}{cc}
\hspace{-1cm}
{\psfig{figure=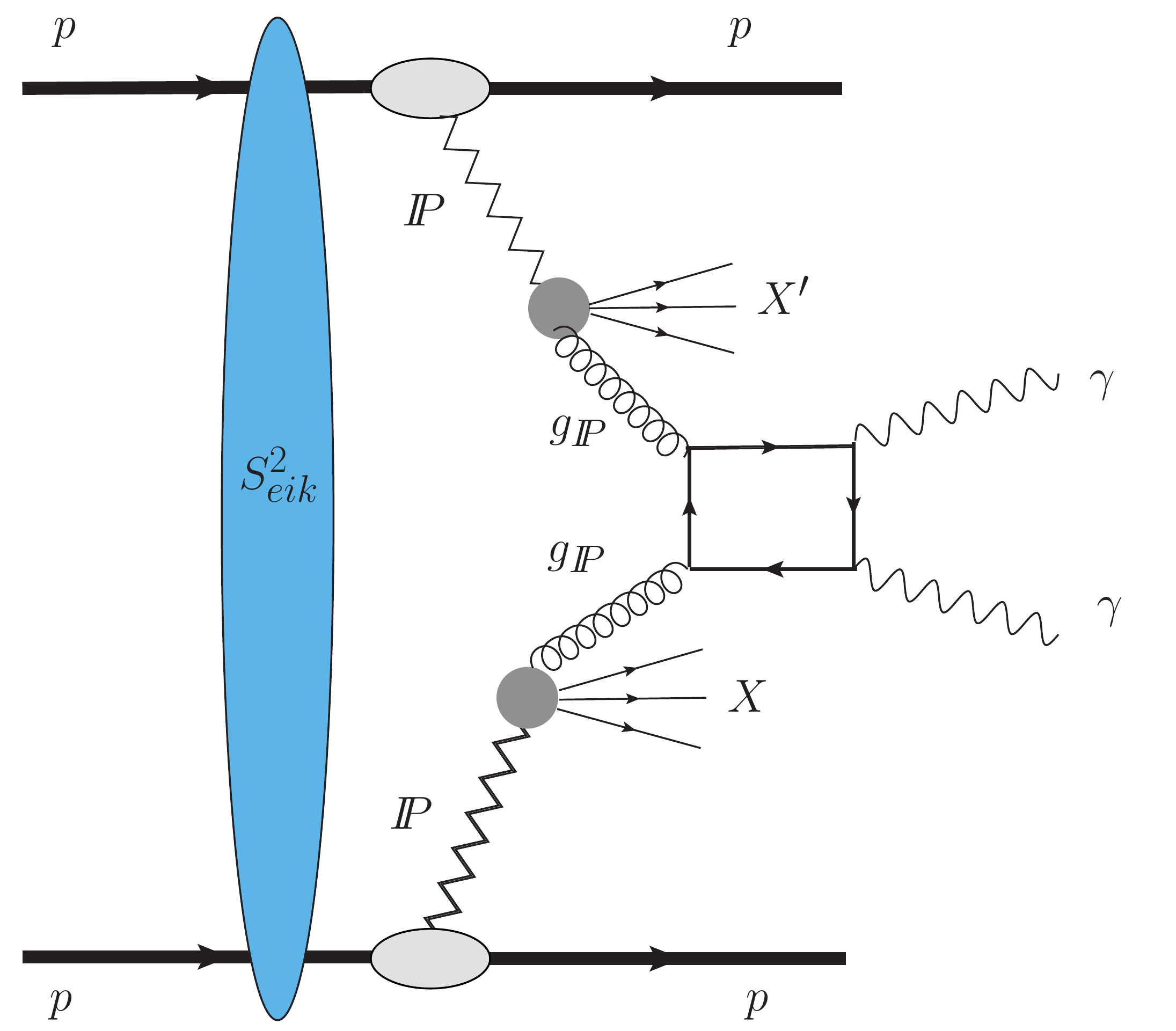,width=5.2cm}}&
{\psfig{figure=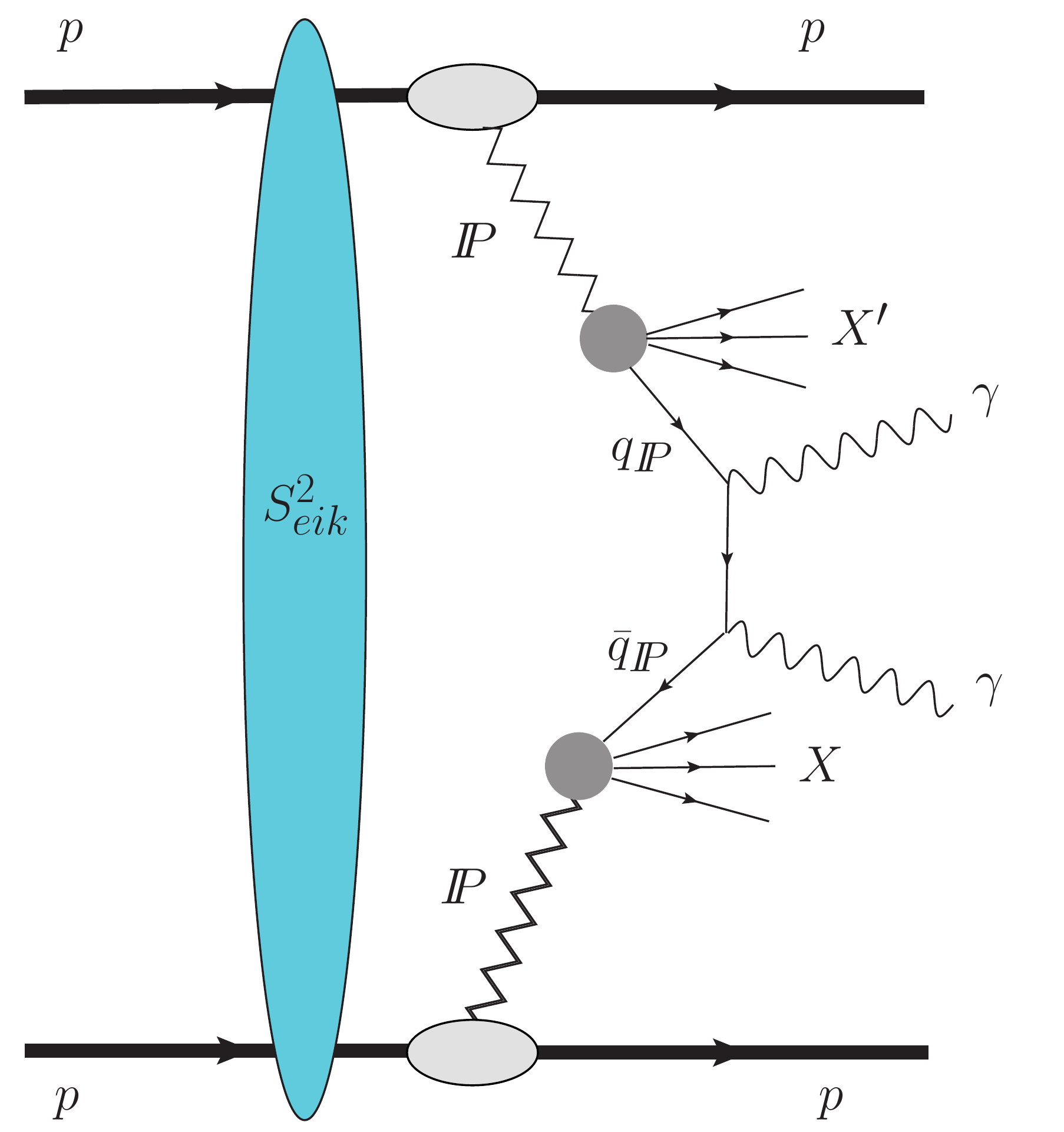,width=4.2cm}} \\
(a) & (b) \\ 
\hspace{-1cm}
{\psfig{figure=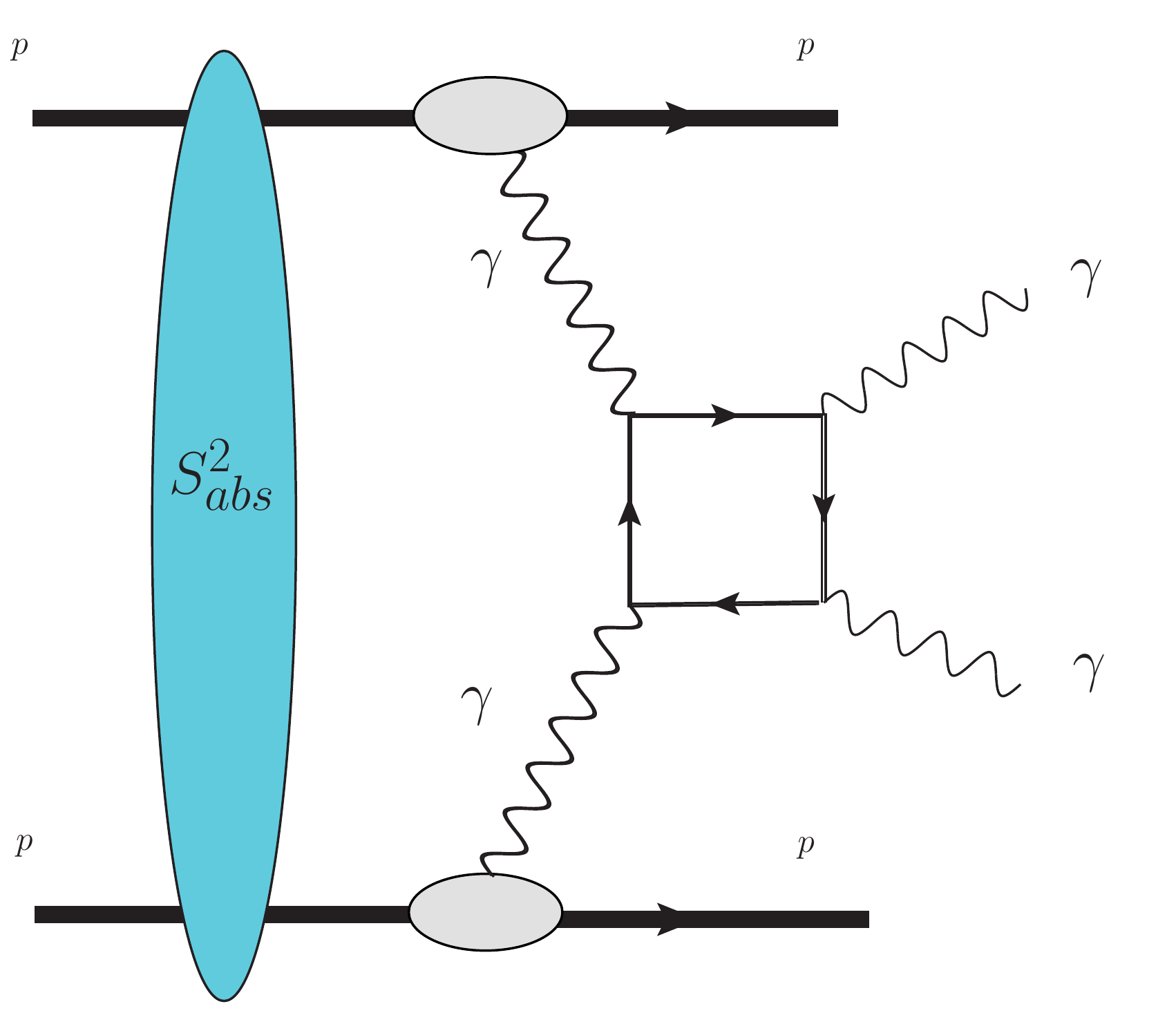,width=5.2cm}} & 
{\psfig{figure=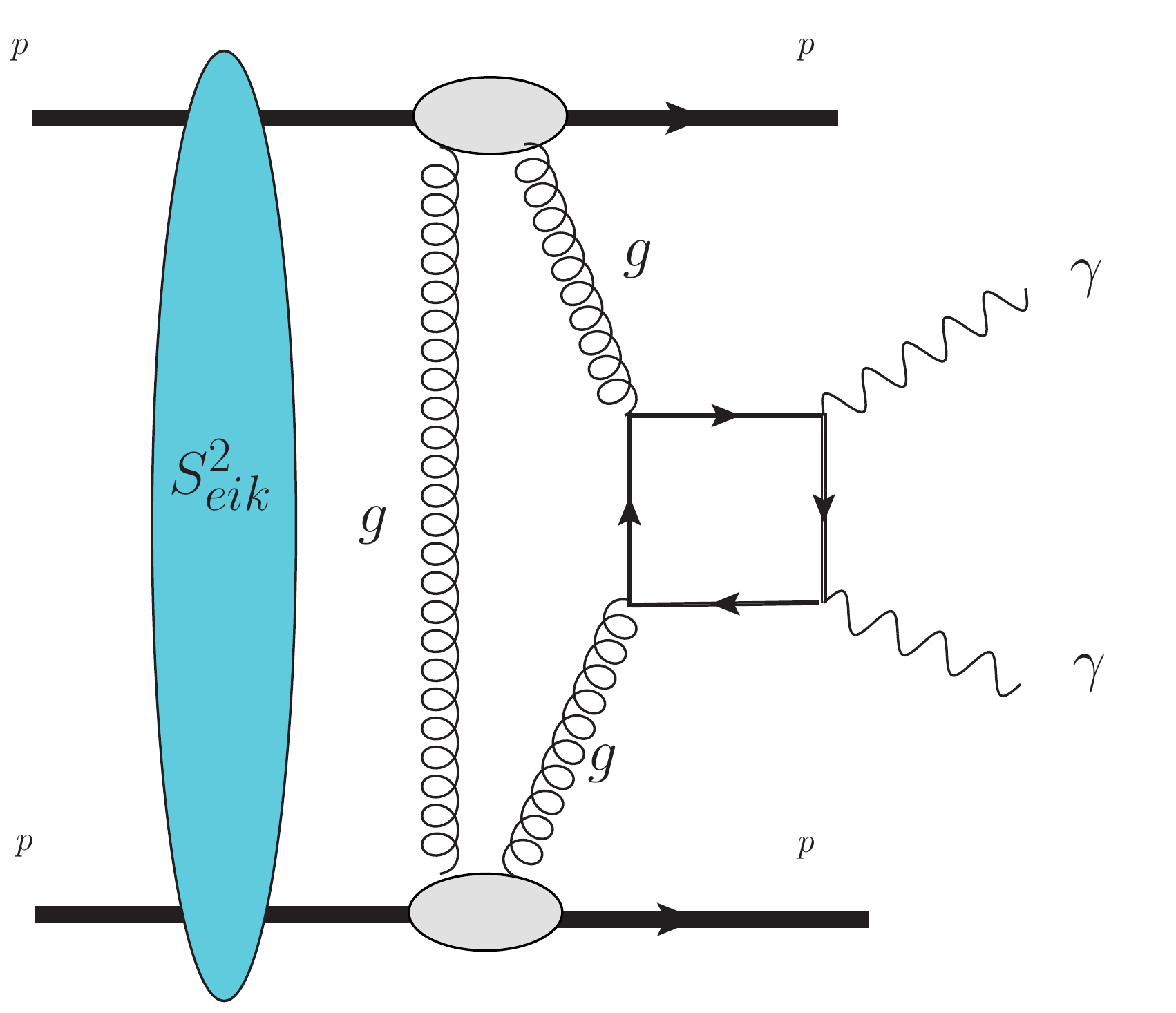,width=5.2cm}} \\ 
(c) & (d) \\ 
\end{tabular}                                                                                                                       
\caption{Diphoton production in $pp$ collisions in  double diffractive processes induced by (a) gluons and (b) quarks of the Pomeron ($\pom$). Backgrounds associated to  the
 (c)  Light -- by -- Light scattering and (d) the central exclusive process induced by gluons (Durham process) are also presented.}
\label{fig:diagram}
\end{figure}

The study of diphoton production in exclusive processes in hadronic collisions became  an active field of research during recent years, strongly motivated by the possibility to observe one of the main consequences of the Quantum Electrodynamics (QED): the 
Light-by-Light (LbL) scattering. Although several attempts were made to detect such rare phenomenon, e.g., the high precision measurements of the electron and muon anomalous magnetic moment~\cite{VanDyck101103,g2collab101103}, direct observations in the laboratory remained challenging until CMS and ATLAS Collaboration have observed, for the first time, the LbL scattering  in ultraperipheral $PbPb$ Collisions~\cite{Aad:2019ock,Sirunyan:2018fhl}. Such collisions are characterized by an impact parameter $b$ greater than the sum of the radius of the colliding nuclei~\cite{upc1,upc2,upc3,upc4,upc5,upc6,upc7,upc8,upc9} and by a photon -- photon luminosity that scales with $Z^4$, where $Z$ is number of protons in the nucleus. As a consequence, in ultraperipheral heavy ion collisions (UPHIC),   the elementary  elastic  $\gamma \gamma \rightarrow \gamma \gamma$ process, which  occurs at one -- loop level at order $\alpha^4$ and have a tiny cross section, is enhanced by a large $Z^4$ ($\approx 45 \times 10^6$) factor. In addition, the contribution of gluon initiated processes can be strongly reduced in nuclear collisions \cite{nos_epjc},  becoming the LbL scattering feasible for the experimental analysis~\cite{gustavo,antoni}. On the other hand, for $pp$ collisions, due the absence of the $Z^4$ enhancement, the diphoton production by gluon initiated processes are expected to significantly contribute and can be dominant in some regions of the phase space (For previous theoretical and experimental studies see, e.g. Refs. \cite{kmr_diphoton,pire,tevatron}). Our goal in this letter is to estimate the contribution of the double diffractive processes, represented in Figs. \ref{fig:diagram} (a) and (b), for the diphoton production in $pp$ collisions at $\sqrt{s} = 13$ TeV. In such reactions, the diphoton system  is generated by the interaction between partons (quarks and gluons)  of the Pomeron ($\pom$), which is a color singlet object inside the proton.
The associated final state  will be characterized by the diphoton system,  two intact protons and  two rapidity gaps, i.e. empty regions  in pseudo-rapidity that separate the intact very forward protons from the $\gamma \gamma$ system. In principle, these events can be separated by tagging the intact protons in the final state using forward detectors, as e.g. the AFP/ATLAS \cite{afp1,afp2} and the CT -- PPS \cite{pps}, and/or by measuring the rapidity gaps. In addition,  to separate the double diffractive processes, we must to control the background associated to the LbL scattering and the exclusive diphoton process, represented in Figs. \ref{fig:diagram} (c) and (d), respectively. In our analysis, we will estimate all these processes using the Forward Physics Monte Carlo (FPMC) \cite{fpmc} and SuperChic event generator \cite{SC3}, taking into account the acceptance of the LHC detectors. In particular, we will  consider the typical set of cuts used by the ATLAS and CMS Collaborations to separate the exclusive events. In addition, we will present, for the first time, the predictions for the diphoton production in double diffractive processes for the kinematical range probed by the LHCb detector. As we will show below, the events associated to double diffractive processes can be separated by imposing a cut on the transverse momentum of the diphoton system, which allow us to investigate the dependence of the predictions on the modeling of the diffractive parton distributions.


Initially, let's present a short review of the main aspects need  to describe  the diphoton production in the double diffractive processes (DDP), represented in Figs. \ref{fig:diagram} (a) and (b). The corresponding cross section can be expressed by
\begin{eqnarray}
\sigma(p p \rightarrow p \otimes X +  \gamma \gamma +  X^{\prime} \otimes p) & = &  \left\{ \int dx_{1} \int dx_{2} \, \left[g^D_{1}(x_{1},\mu^2) \cdot g^D_{2}(x_{2},\mu^2) \cdot \hat{\sigma}(g g \rightarrow \gamma \gamma)\right.\right.  \nonumber \\
& + & \left. \left. \, [q^D_{1}(x_{1},\mu^2) \cdot \bar{q}^D_{2}(x_{2},\mu^2) + \bar{q}^D_{1}(x_{1},\mu^2) \cdot {q}^D_{2}(x_{2},\mu^2)]  \cdot \hat{\sigma}(q \bar{q} \rightarrow \gamma \gamma)\right]\right\} \,\,,
\label{pompom}
\end{eqnarray}
where $g^D_i (x_i,\mu^2)$, $q^D_i (x_i,\mu^2)$ and $\bar{q}^D_i (x_i,\mu^2)$ are the diffractive gluon, quark and antiquark densities of the proton $i$ with a momentum fraction $x_i$. The  parton distributions have its evolution in the hard scale $\mu^2$ given by the DGLAP evolution equations and should be determined from events with a rapidity gap or a intact hadron. In the  Resolved Pomeron model \cite{IS} the diffractive parton distributions  are expressed in terms of parton distributions in the {pomeron} and a Regge parametrization of the flux factor describing the {Pomeron} emission by the hadron. In particular, the diffractive gluon distribution can be expressed   as follows 
\begin{eqnarray}
{ g^D_{p}(x,\mu^2)}= { \int_x^1 \frac{d\xi}{\xi} f^{p}_{\pom}(\xi) ~g_{\pom}\left(\frac{x}{\xi}, \mu^2\right)}   \,\,,
\label{difquark:proton}
\end{eqnarray}
where $\xi$ is the momentum fraction of the proton carried by the Pomeron, $f^{p}_{\pom}(\xi)$ stands for the associated flux distributions in the proton and  
$g_{\pom}(\beta \equiv {x}/\xi , \mu^2)$ is the Pomeron  gluon distribution (A similar definition is valid for the diffractive quark distribution). Furthermore,  $\beta $ is the momentum fraction carried by the gluon inside the Pomeron. It is useful to  assume that the  \,{Pomeron} flux is given by
\begin{eqnarray}
f^{p}_{\pom}(\xi)= 
\int_{t_{\rm min}}^{t_{\rm max}} dt \, \frac{A_{\pom} \, e^{B_{\pom} t}}{\xi^{2\alpha_{\pom} (t)-1}},
\label{fluxpom:proton}
\end{eqnarray}
where $t_{\rm min}$, $t_{\rm max}$ are kinematic boundaries. The flux factors are motivated by Regge theory, where the \,{Pomeron} trajectories are assumed to be linear, $\alpha_{\pom} (t)= \alpha_{\pom}(0) + \alpha_{\pom}^\prime t$, and the parameters $B_{\pom}$, $\alpha_{\pom}^\prime$  and their uncertainties are obtained from fits to the  data. In our analysis, we will consider different parametrizations for $g_{\pom}(\beta , \mu^2)$ and $q_{\pom}(\beta , \mu^2)$  in order to investigate   the sensitive of the predictions on the description of the Pomeron structure. In order to derive realistic predictions for the diphoton production in double diffractive process, we need to take into account of the nonperturbative effects associated to soft interactions which imply the breakdown of the collinear factorization \cite{collins} and 
lead to an extra production of particles that destroy the rapidity gaps in the final state \cite{bjorken}. The treatment of these soft survival corrections is still strongly model dependent (recent reviews can be found in Refs. \cite{durham,telaviv}). In our analysis,  we will assume that the hard process occurs on a short enough timescale such that the physics that generate the additional particles can be factorized and accounted by an overall factor. As a consequence, the soft survival effects can be included in the calculation by multiplying the cross section by  a global factor $S^2_{eik}$ (denoted eikonal factor in  Fig. \ref{fig:diagram}). It is important to emphasize that the validity of this assumption  is still an open question and should be considered a first approximation for this difficult problem. As in Refs. \cite{kkmr,nos_Dijet,nos_dimuons}, we will assume that $S^2_{eik} = 0.03$ for $pp$ collisions at $\sqrt{s} = 13$ TeV.

In what follows, we will estimate the diphoton production in double diffractive processes using  the Forward Physics Monte Carlo (FPMC) \cite{fpmc}, which allow us to estimate the associated cross sections and distributions taking into account of the detector acceptances. In this event generator, the hard matrix elements are treated by interfacing FPMC with HERWIG  v6.5 \cite{herwig} which includes perturbative parton showering followed by the hadronization. In order to estimate the impact of the Pomeron structure in the predictions, we will consider four different parametrizations for the diffractive parton distributions, which are based on different assumptions for the $\beta$ -- behavior  and have been obtained using distinct sets of data ~\cite{H1diff,Royon:2006by}. For the calculation of the LbL scattering  and exclusive $g g \rightarrow \gamma\gamma$ production (Durham model) it is employed a dedicated event generator for exclusive processes: \superchic 3~\cite{SC3}. The soft survival effects $S^2_{{eik}}$ have been include in our calculations of the exclusive process assuming the model 4 implemented in the SuperChic3. For the LbL scattering, we also will consider the survival factor included in this event generator. We have checked that for this process the impact of the soft corrections is small, with  $S^2_{{eik}} \approx 1$. For the proton tagging at the LHC, a region of $0.015\leq \xi \leq 0.15$ for both protons is chosen for a center of mass energy of 13 TeV considering the standard acceptance in the central detectors.
For the final state selection we use \hepmc 2~\cite{hepmc} with a plugin called \heppdt \footnote{Available at http://lcgapp.cern.ch/project/simu/HepPDT/} which has been designed to be used by any Monte Carlo particle generator or decay package. \heppdt~ has the function to store particle information such as charge and nominal mass in a table which is
accessed by a particle ID number. The particle ID number is defined according to the Particle Data Group’s Monte Carlo numbering scheme~\cite{pdg}.
The final analysis and distributions are done with \analysisroot~\cite{root}.

Two distinct configurations of experimental offline cuts are considered: one refers to a typical central detector as ATLAS and CMS, and other for a forward detector such as LHCb.  The protons are assumed to be intact in the interaction, allowing to measure the central mass, $m_{X}=\sqrt{\xi_{1}\xi_{2}s}$, where $\xi_{1,2}$ is the proton fraction momentum loss given by $1-(p_{Z_{1,2}}/6500)$ and $\sqrt{s}$ is the center of mass energy. This quantity can be studied in dedicated very forward detectors AFP and CT-PPS \cite{afp1,afp2,pps} installed around the ATLAS and CMS interaction points. On the other side, the LHCb experiment is able to veto particles using forward shower counters (HERSCHEL) within the acceptance $8.0 < |\eta| < 5.5$~\cite{herschel}.

\begin{center}
\begin{table}[t]
\begin{tabular}{c|c|c|c|c|c|c}
\hline 
{\bf $pp$ collisions at $\sqrt{s}$ = 13 TeV} & LbL & Durham & DDP (H1-FitA) & DDP (H1-FitB) & DDP (ZEUS) & DDP (H1-ZEUS) \tabularnewline
\hline 
Total Cross section {[}pb{]} & 1.68 & 305.0&  95.0 & 150.1  & 56.4 & 98.9 
\tabularnewline
\hline
\end{tabular}
\caption{Predictions for the diphoton production in double diffractive processes considering different modeling of the Pomeron structure. For comparison the predictions associated to the LbL scattering and exclusive (Durham) process are also presented. Results at the generation level.}
\label{tab:gen}
\end{table}
\end{center}

\begin{center}
 \begin{figure}[t]
 \includegraphics[width=0.32\textwidth]{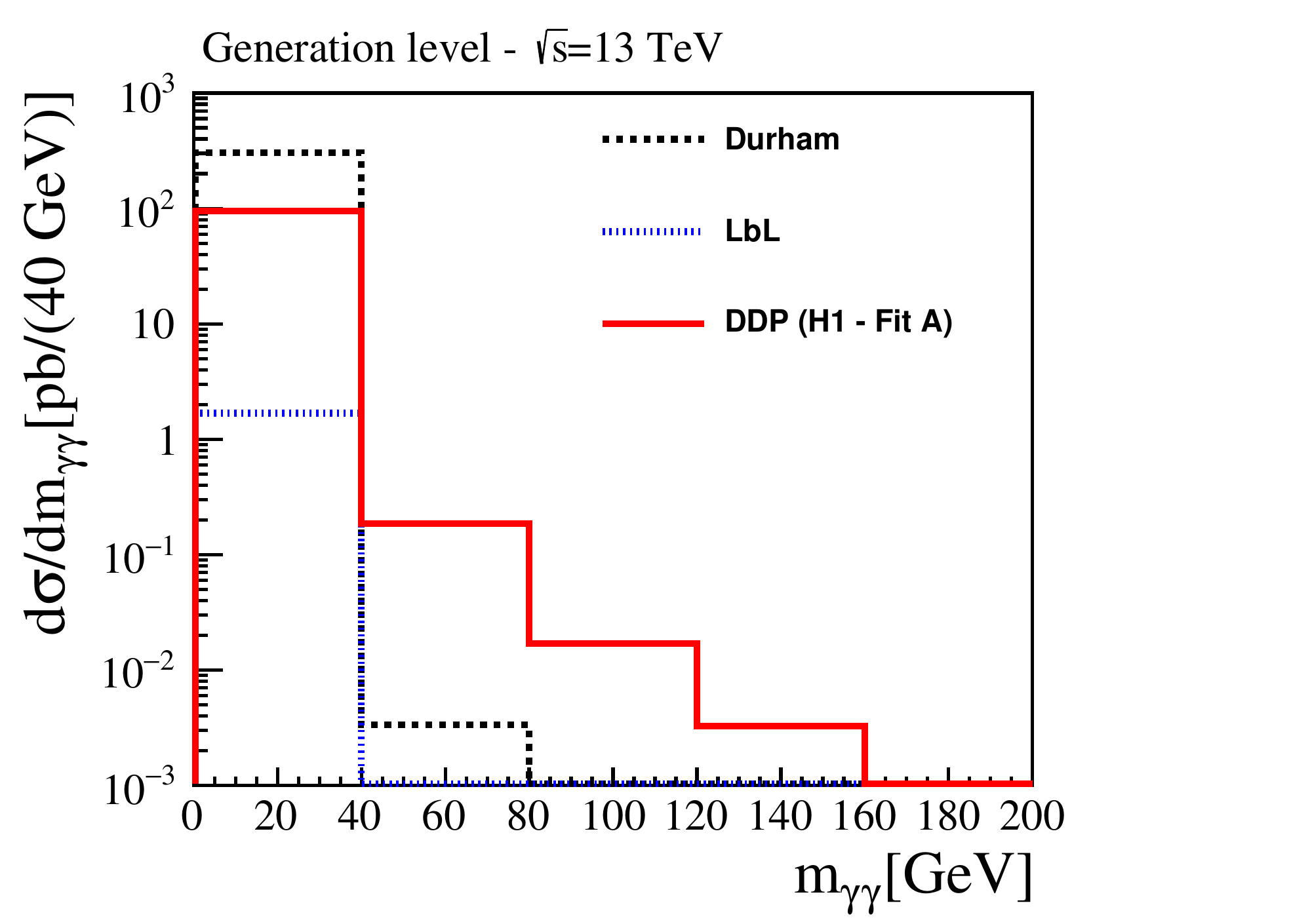}
 \includegraphics[width=0.32\textwidth]{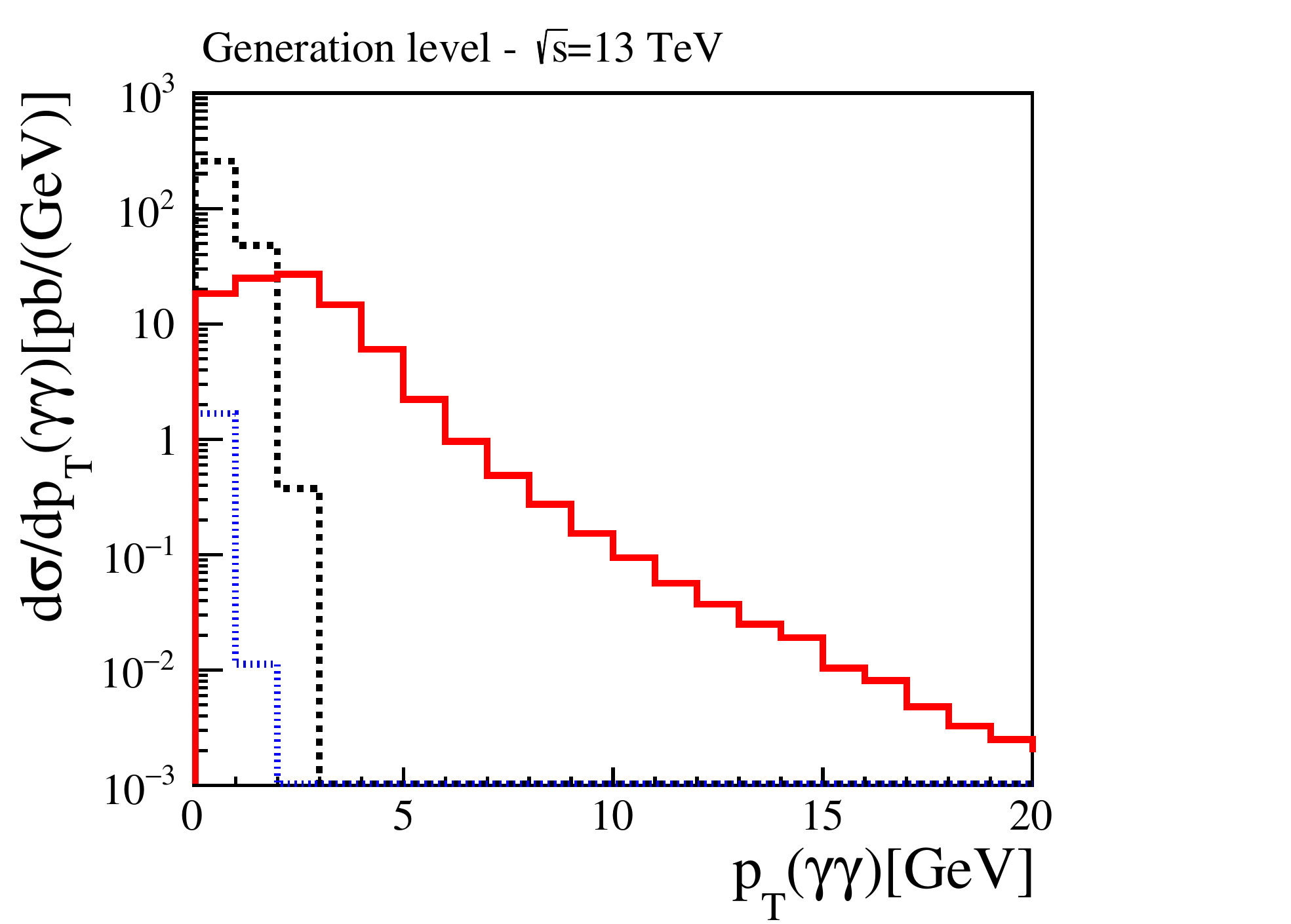}
 \includegraphics[width=0.32\textwidth]{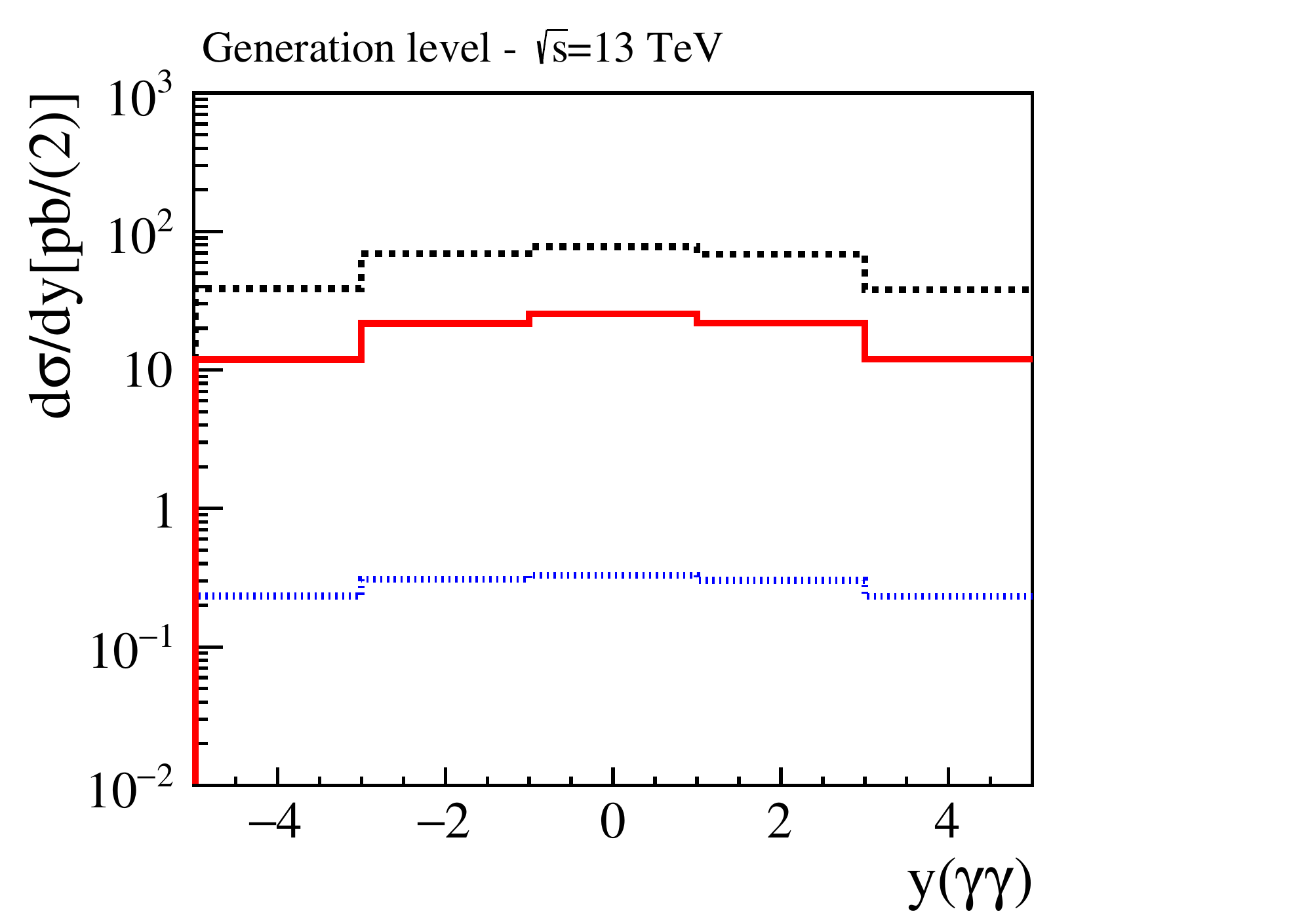}
 \caption{Predictions for the invariant mass $m_{\gamma\gamma}$, transverse momentum $p_{T}(\gamma\gamma)$  and rapidity $y_{\gamma \gamma}$ distributions of the diphoton system produced in $pp$ collisions at the LHC. Results obtained at the generation level, without the inclusion of experimental cuts.}
 \label{fig:generation}
 \end{figure}
 \end{center}

Initially, in Table \ref{tab:gen} we  present our results for the  cross sections associated to the different channels, obtained at the generation level, without the inclusion of any selection in the events. The predictions for the double diffractive processes are presented considering four distinct parametrizations for the diffractive parton distributions. 
We have that the gluon -- induced processes (Durham and DDP) are dominant, with the DDP predictions being strongly dependent on the Pomeron structure.
In Fig. 
\ref{fig:generation} we present our results for the invariant mass $m_{\gamma \gamma}$, transverse momentum $p_T(\gamma \gamma)$  and  rapidity $y (\gamma \gamma)$ distributions of the diphoton system. The DDP and Durham predictions are larger than the LbL one in the kinematical range considered. Such result contrast with that presented in Ref. \cite{nos_epjc} for $PbPb$ collisions, where the LbL scattering is dominant due to the $Z^4$ enhancement. For   $m_{\gamma \gamma} \le 40$ GeV and transverse momentum $p_T(\gamma \gamma) \le 2$ GeV, the Durham process dominates. On the other hand, for larger values of $m_{\gamma \gamma}$ and $p_T(\gamma \gamma)$, the main contribution for the diphoton production comes from double diffractive processes.

In order to obtain realistic estimates of the diphoton production in $pp$ collisions, which can be compared with the future experimental data, we will include in our analysis the experimental cuts  that are expected to be feasible in the next run of the LHC. The selection criteria implemented in our analysis of double diffractive and exclusive diphoton processes are the following:   
\begin{itemize}
\item For a central detector: We will select events in which $m(\gamma\gamma)$ > 5 GeV and $E_{T}(\gamma,\gamma)$ > 2 GeV, where $E_{T}$ is the transverse energy of the photons. Moreover, we will impose a cut on the transverse momentum of the diphoton system  ($p_{T}(\gamma\gamma) > 3$ GeV). Finally, we only will select events where photons are produced in the rapidity range $|\eta(\gamma^{1},\gamma^{2})| < 2.5$ and the proton fraction momentum loss window $0.015\leq \xi_{1,2}\leq 0.15$ which corresponds to the central mass $m_X$ larger than 195 GeV, the kinematical range covered by the forward detectors AFP and CT --  PPS \cite{afp1,afp2,pps}.

\item For a forward detector: We will select events in which $m(\gamma\gamma)$ > 1 GeV and $p_{T}(\gamma,\gamma)$ > 0.2 GeV, where $p_T$ is the transverse momentum of the photons. An additional cut is applied on transverse momentum of the diphoton system ($p_{T}(\gamma\gamma)$ > 3 GeV). Finally, we will select only events where photons are produced in the rapidity range $2.0 < |\eta(\gamma^{1},\gamma^{2})| < 4.5$, not allowing particles with $p_T > 0.5$ GeV in the range $8.0 < |\eta| < 5.5$, corresponding to the HERSCHEL selection in the LHCb.
\end{itemize}

\begin{center}
\begin{table}[t]
\begin{tabular}{|c|c|c|c|c|c|c|}
\hline 
{\bf $pp$ collisions at $\sqrt{s}$ = 13 TeV} & LbL & Durham & DDP(H1-FitA) & DDP(H1-FitB) & DDP(ZEUS) & DDP(H1-ZEUS) \tabularnewline
\hline 
Total Cross Section {[}pb{]} & 1.68 & 305.0 & 95.0 & 150.1 & 56.4& 98.9  
\tabularnewline
\hline 
$m_{\gamma\gamma}> 5\:\rm{GeV}, E_{T}(\gamma,\gamma)>2\:\rm{GeV}$&0.029   & 6.9    & 31.6 &39.3 &18.5 & 13.6     \tabularnewline
\hline 
$p_{T}(\gamma\gamma)> 3$  GeV &0.0 & 0.0  &11.1  & 14.3 &6.2& 4.3    \tabularnewline
\hline 
 $|\eta(\gamma,\gamma)|<2.5$& 0.0 & 0.0 & 6.3  & 8.5 &3.6 & 2.8 \tabularnewline
\hline
 \hline 
 $ 0.015 \leq \xi_{1,2} \leq 0.15 $& 0.0 & 0.0  & 4.0  &5.7 & 2.3 & 2.0 \tabularnewline
\hline
\hline 
\end{tabular}
\caption{Predictions for the double diffractive diphoton  cross sections after the inclusion of the exclusivity cuts for a typical central detector. For comparison the predictions associated to the LbL scattering and exclusive (Durham) process are also presented.}
\label{tab:central}
\end{table}
\end{center}

The impact of each of these cuts on the total cross sections is summarized in Tables \ref{tab:central} and \ref{tab:forward} for the central and forward detectors, respectively. 
Our results indicate that the inclusion of all cuts fully suppress the contribution of the LbL scattering and exclusive  process for the diphoton production.  We have that these contributions are completely removed by the cut on the transverse momentum of the diphoton system ($p_{T}(\gamma\gamma) \ge 3$  GeV). Such result agrees with those derived in Refs. \cite{nos_epjc,nos_dimuons} and is associated to the fact that that in the double diffractive production the transverse momentum of the gluons inside the Pomeron, which interact to generate the diphoton, can be large. Moreover, it is dependent on the modeling of the diffractive parton distributions. In contrast, in exclusive processes, the initial momentum of the incident particles is restricted by the Pomeron - proton and Photon - proton vertexes, which exponentially suppress larger values of momentum. 
Therefore, the events after cuts are a clean probe of the diphoton production in double diffractive processes.

\begin{center}
 \begin{figure}[t]
 \includegraphics[width=0.32\textwidth]{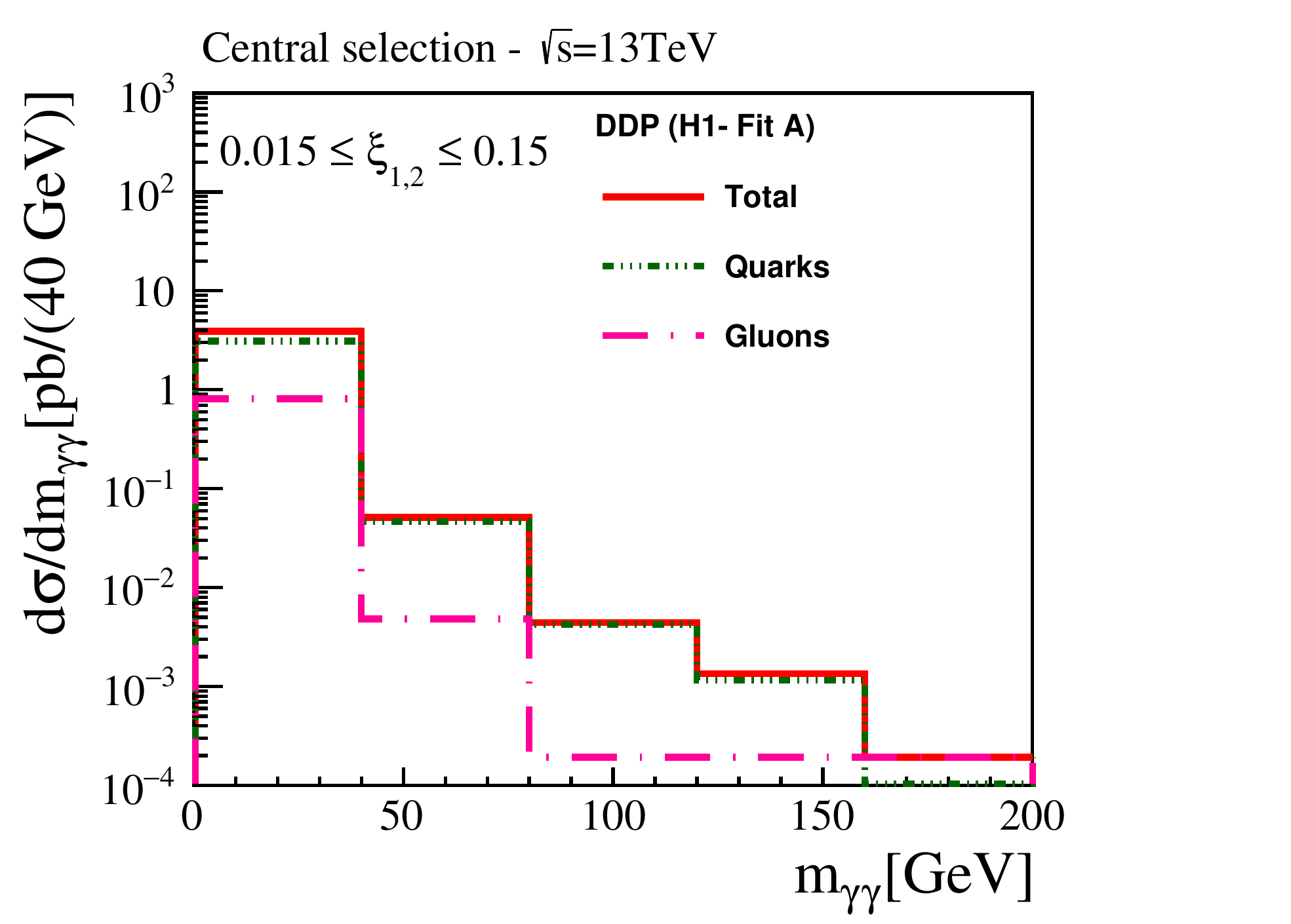}
  \includegraphics[width=0.32\textwidth]{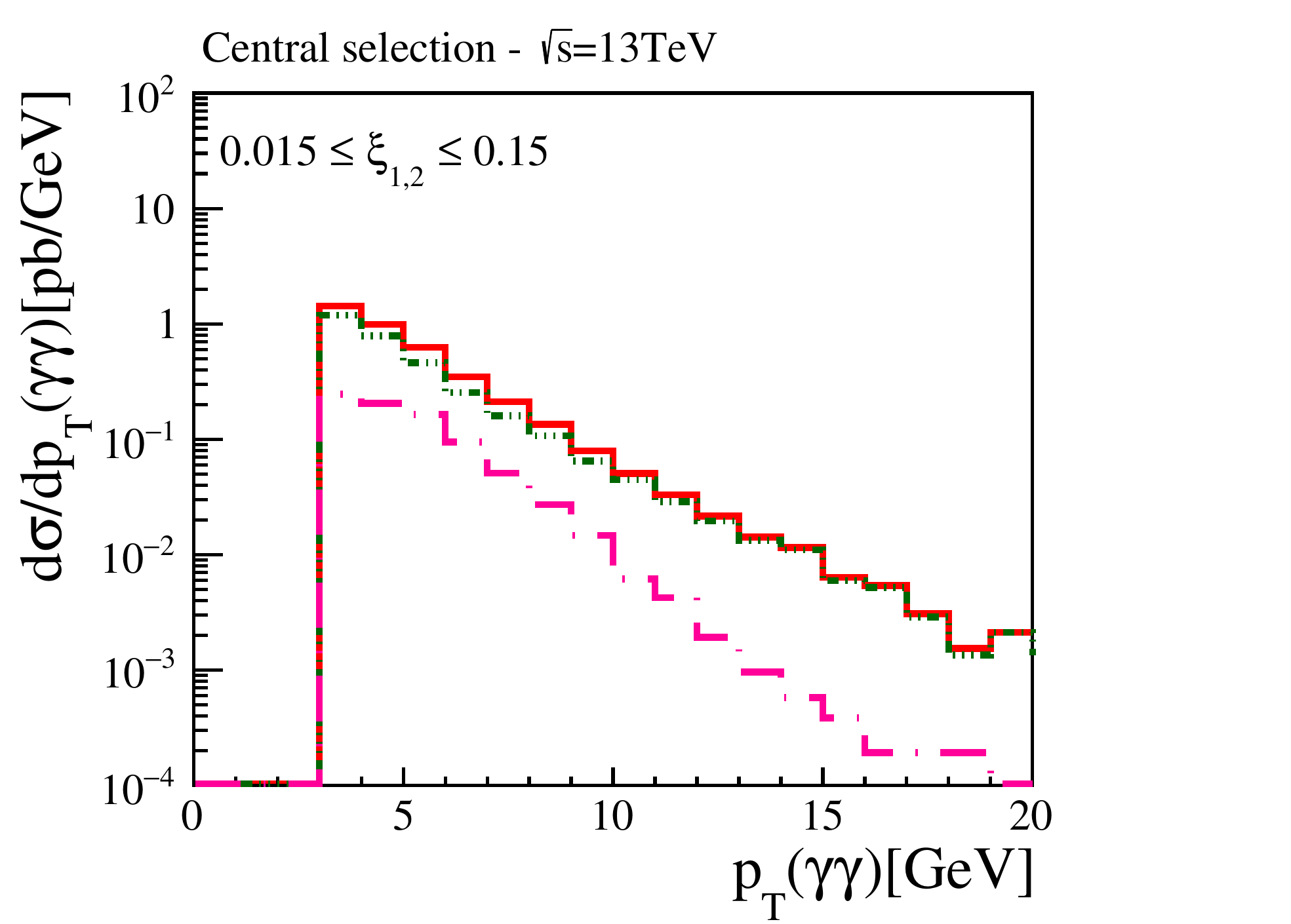}
  \includegraphics[width=0.32\textwidth]{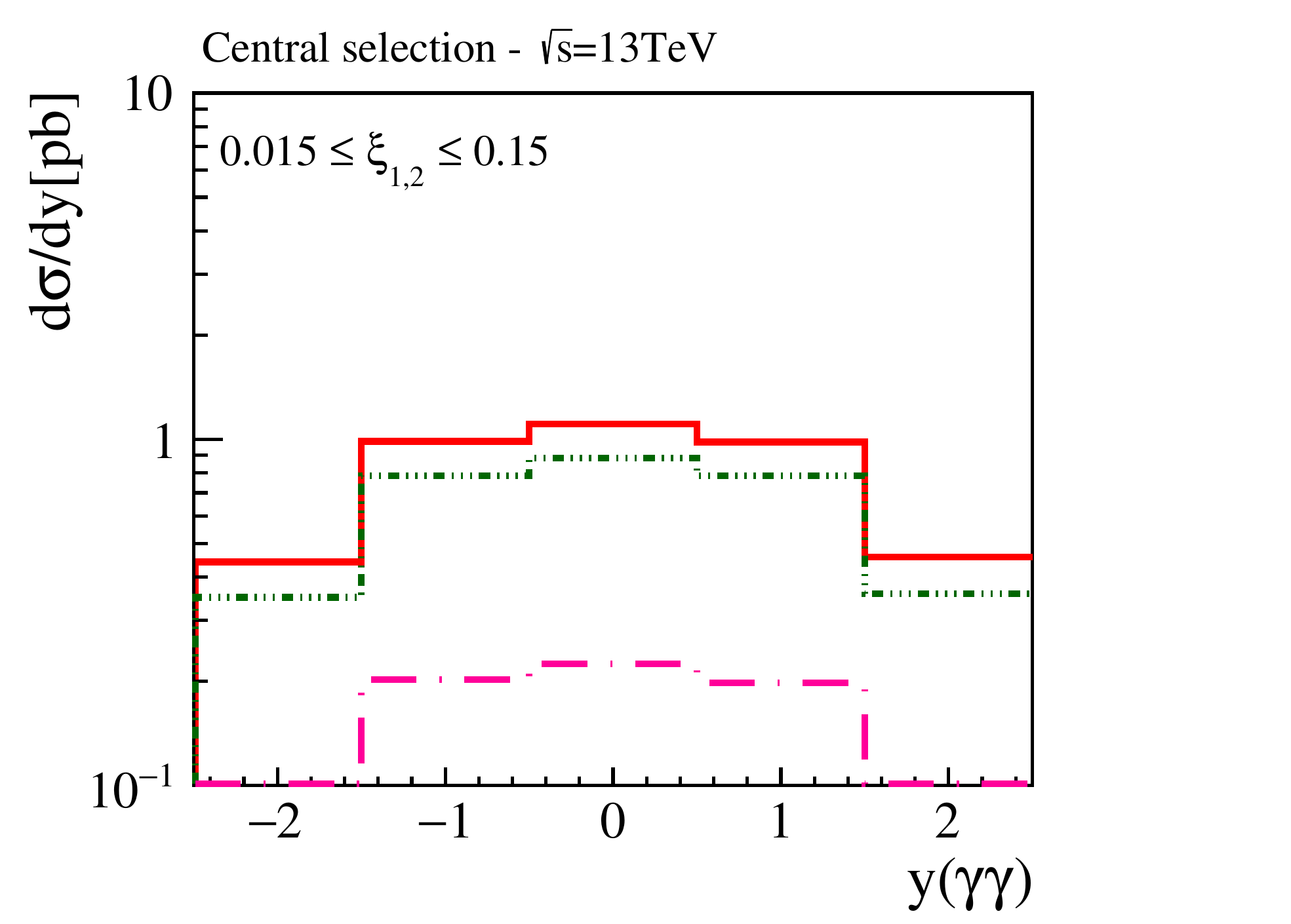}
\includegraphics[width=0.32\textwidth]{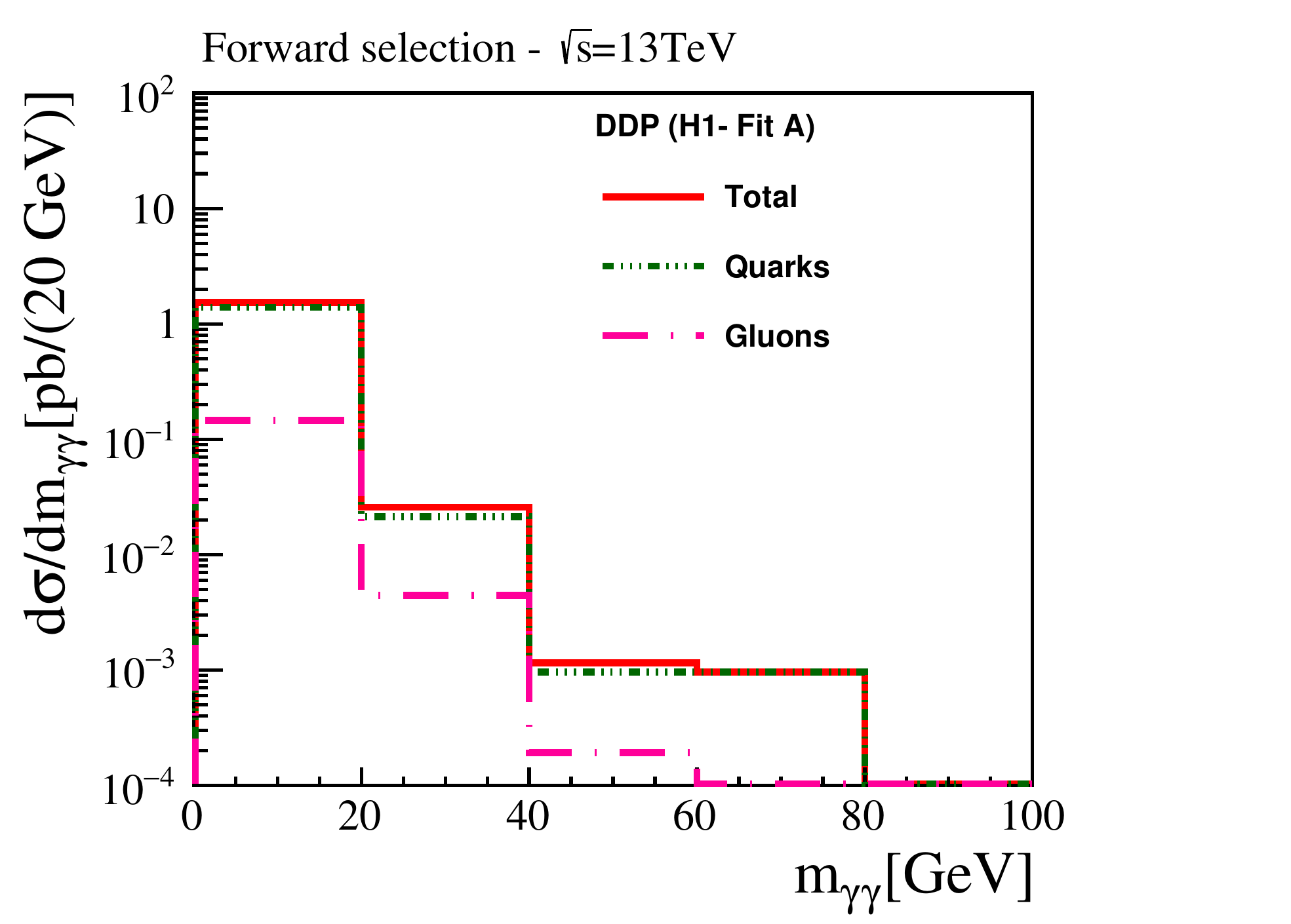}
  \includegraphics[width=0.32\textwidth]{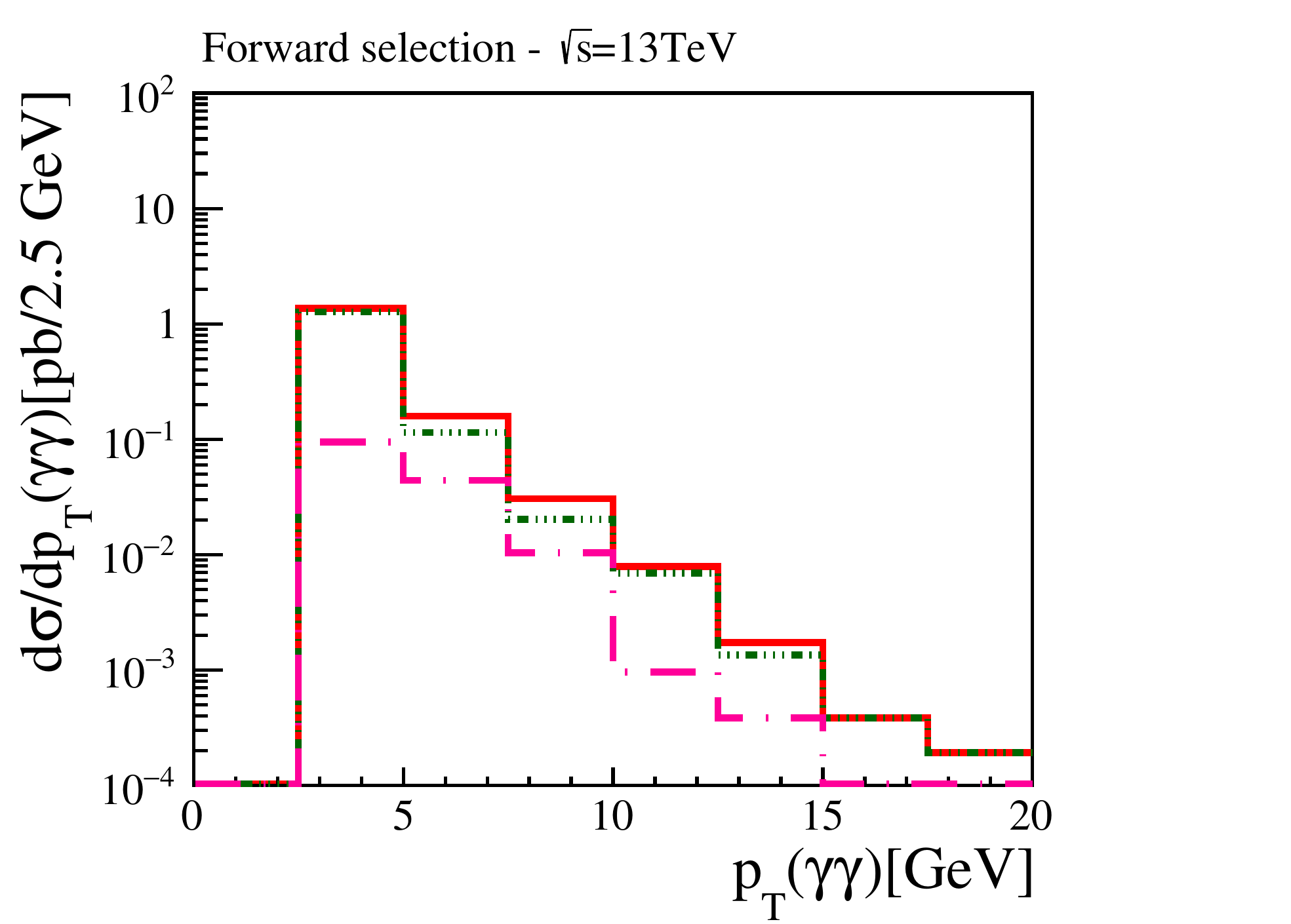}
 \includegraphics[width=0.32\textwidth]{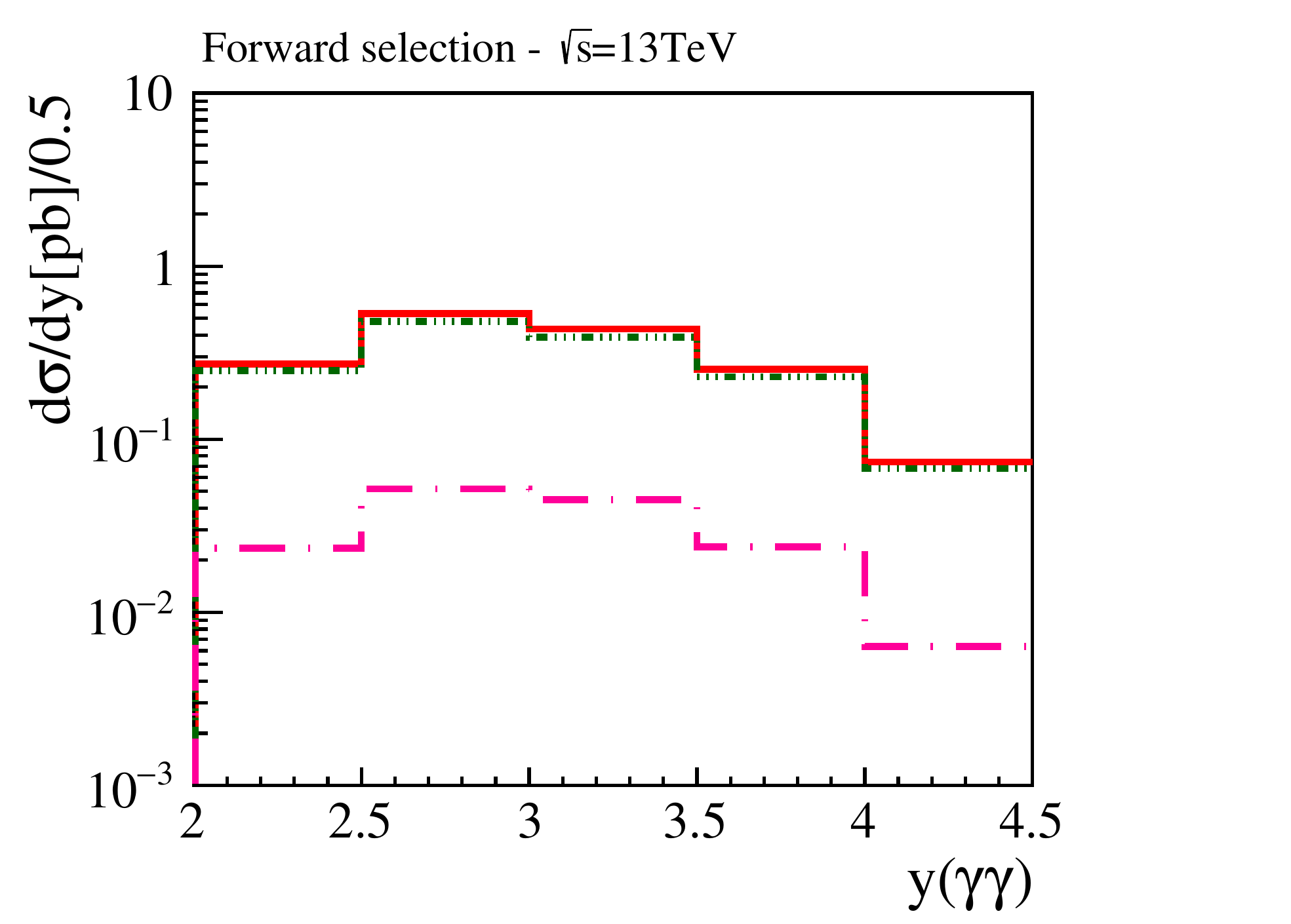}
  \caption{Comparison between the predictions associated to the 
  $q \bar{q} \rightarrow \gamma \gamma$ (denoted quarks) and $g g \rightarrow \gamma \gamma$ (denoted gluons) subprocesses for the  invariant mass $m_{\gamma \gamma}$, transverse momentum $p_T({\gamma \gamma})$  and rapidity $y ({\gamma \gamma})$ distributions of the diphoton system produced in $pp$ collisions at the LHC. Results for a central (forward) detector are presented in the upper (lower) panels.  }
\label{fig:central_gluons}
\end{figure}
\end{center}

\begin{center}
 \begin{figure}[t]
  \includegraphics[width=0.32\textwidth]{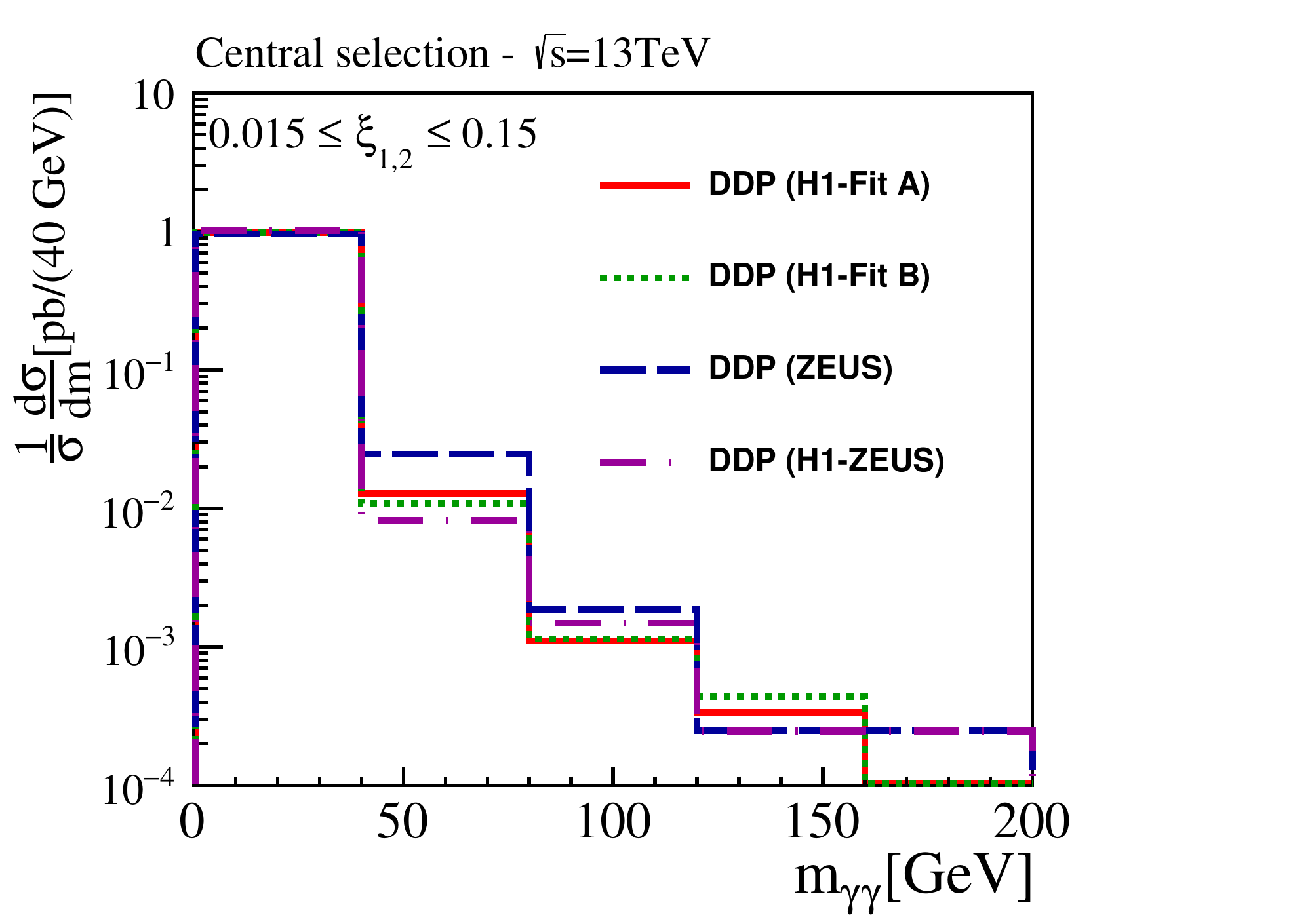}
 \includegraphics[width=0.32\textwidth]{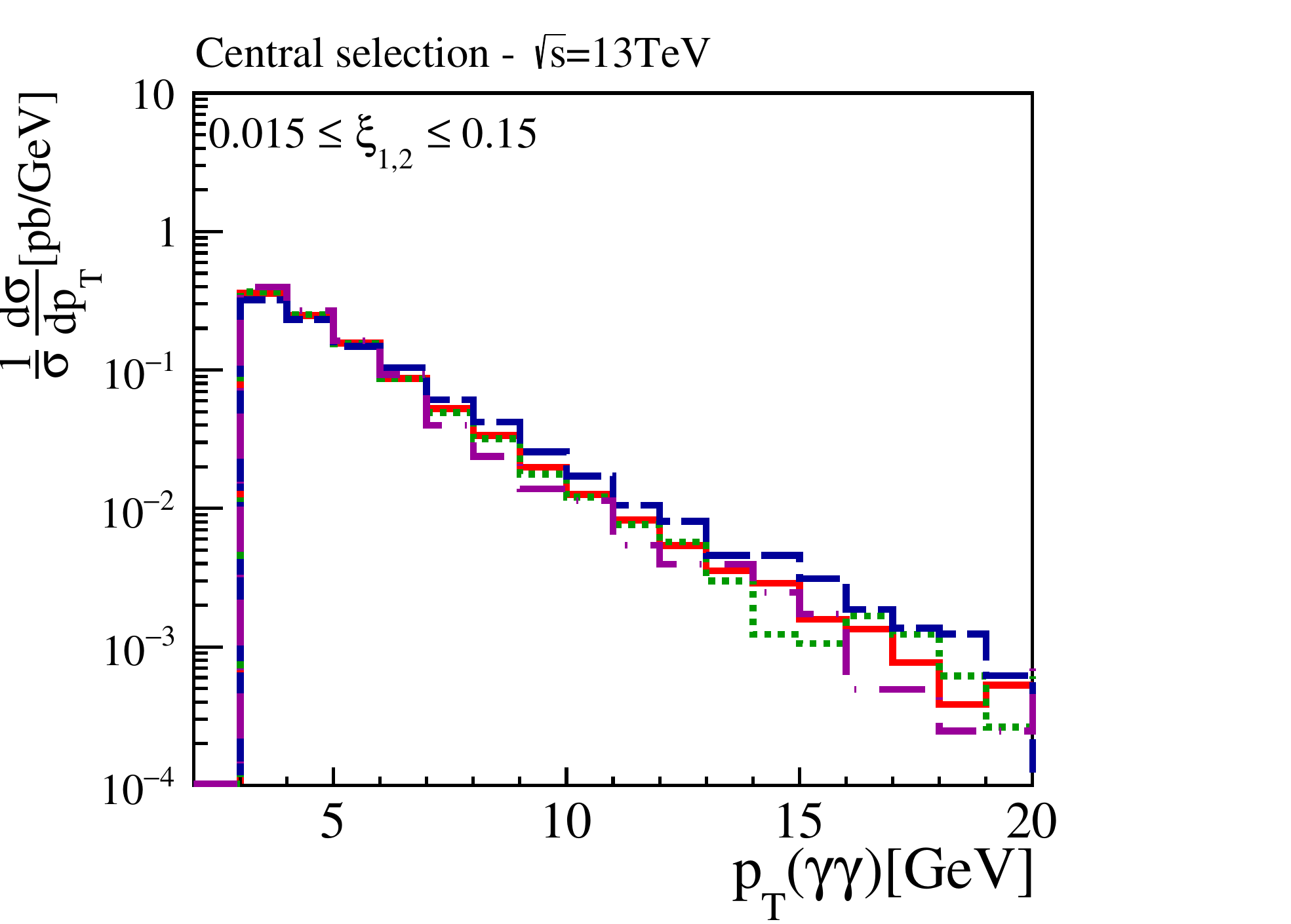}
 \includegraphics[width=0.32\textwidth]{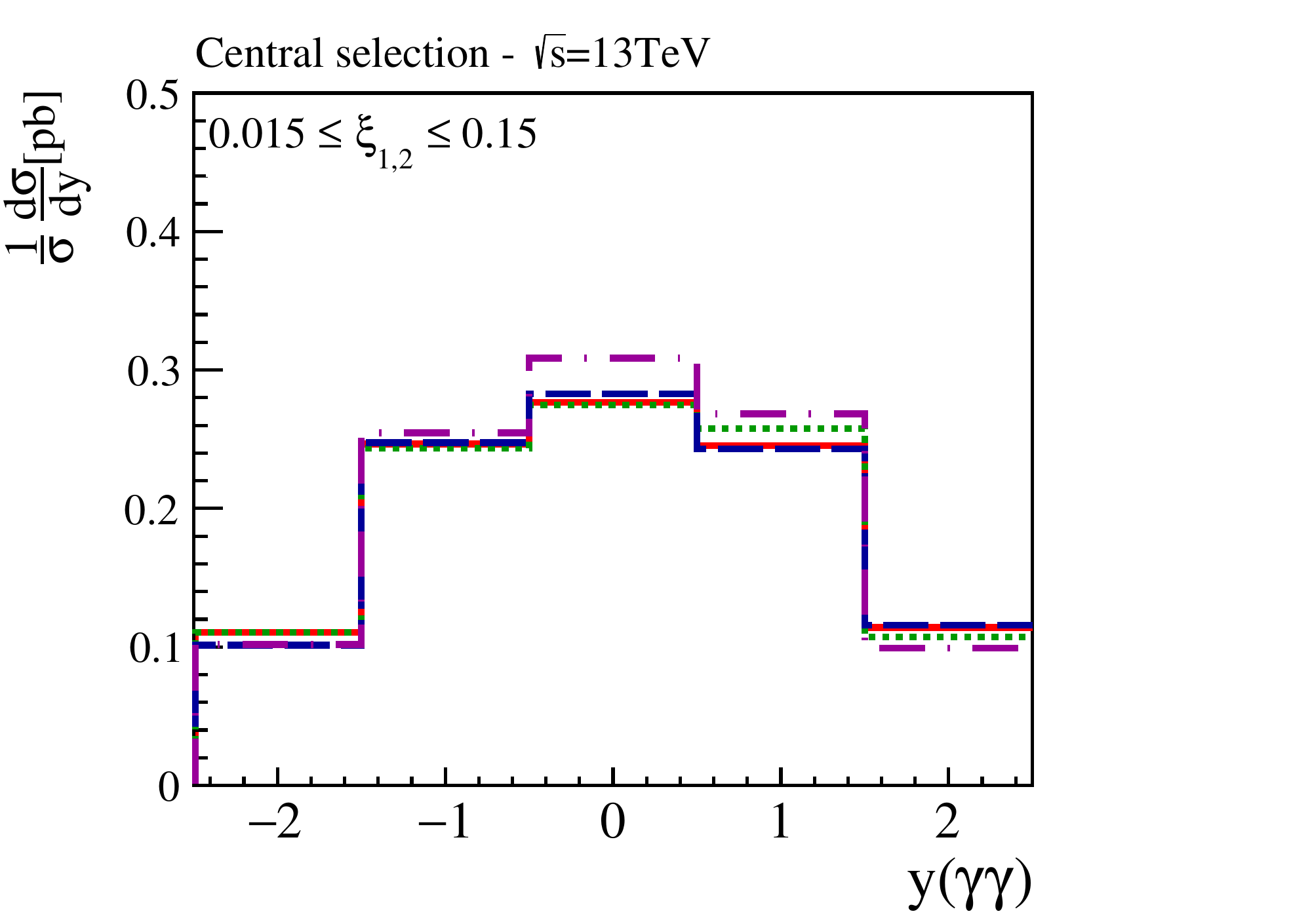}
  \includegraphics[width=0.32\textwidth]{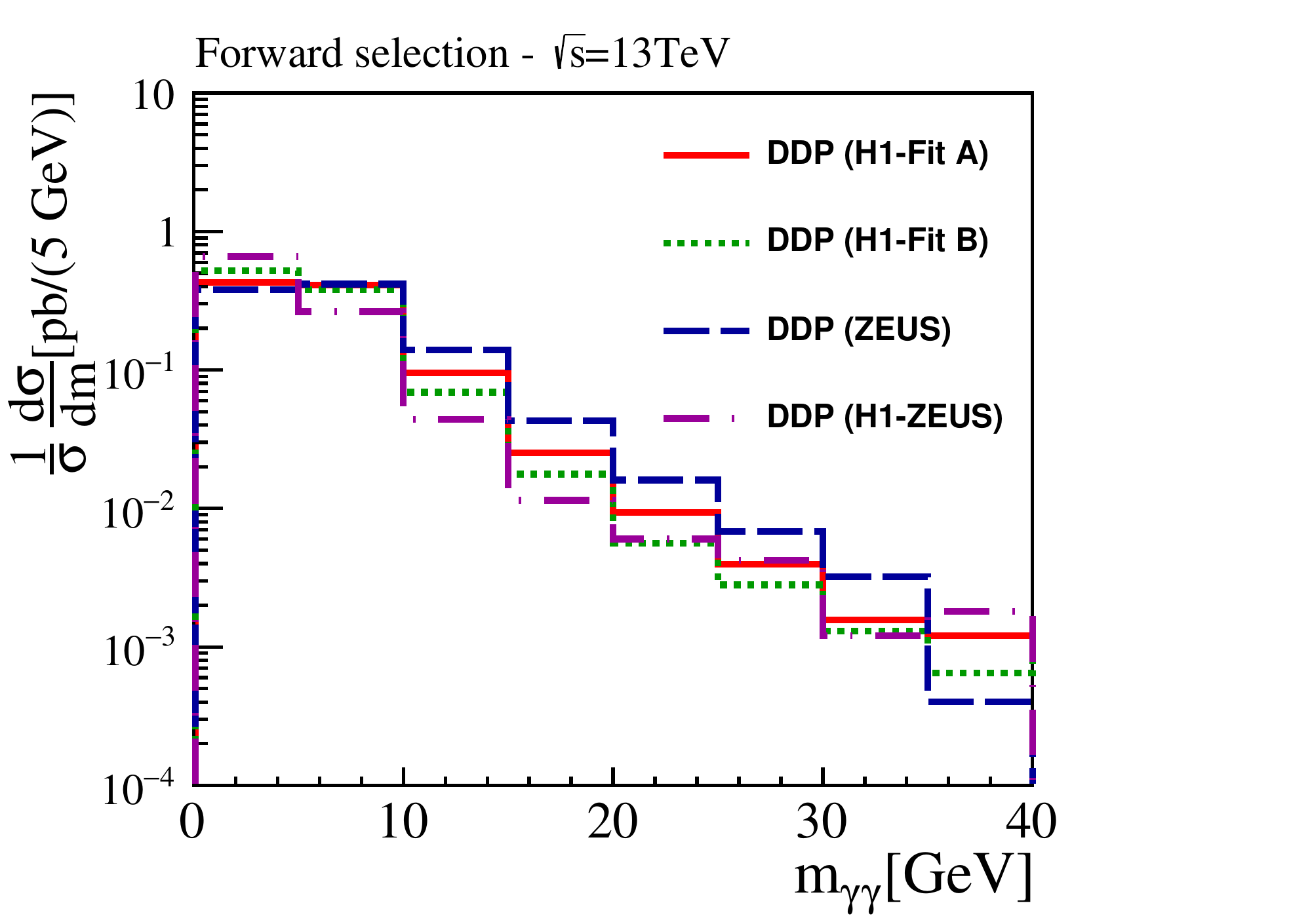}
\includegraphics[width=0.32\textwidth]{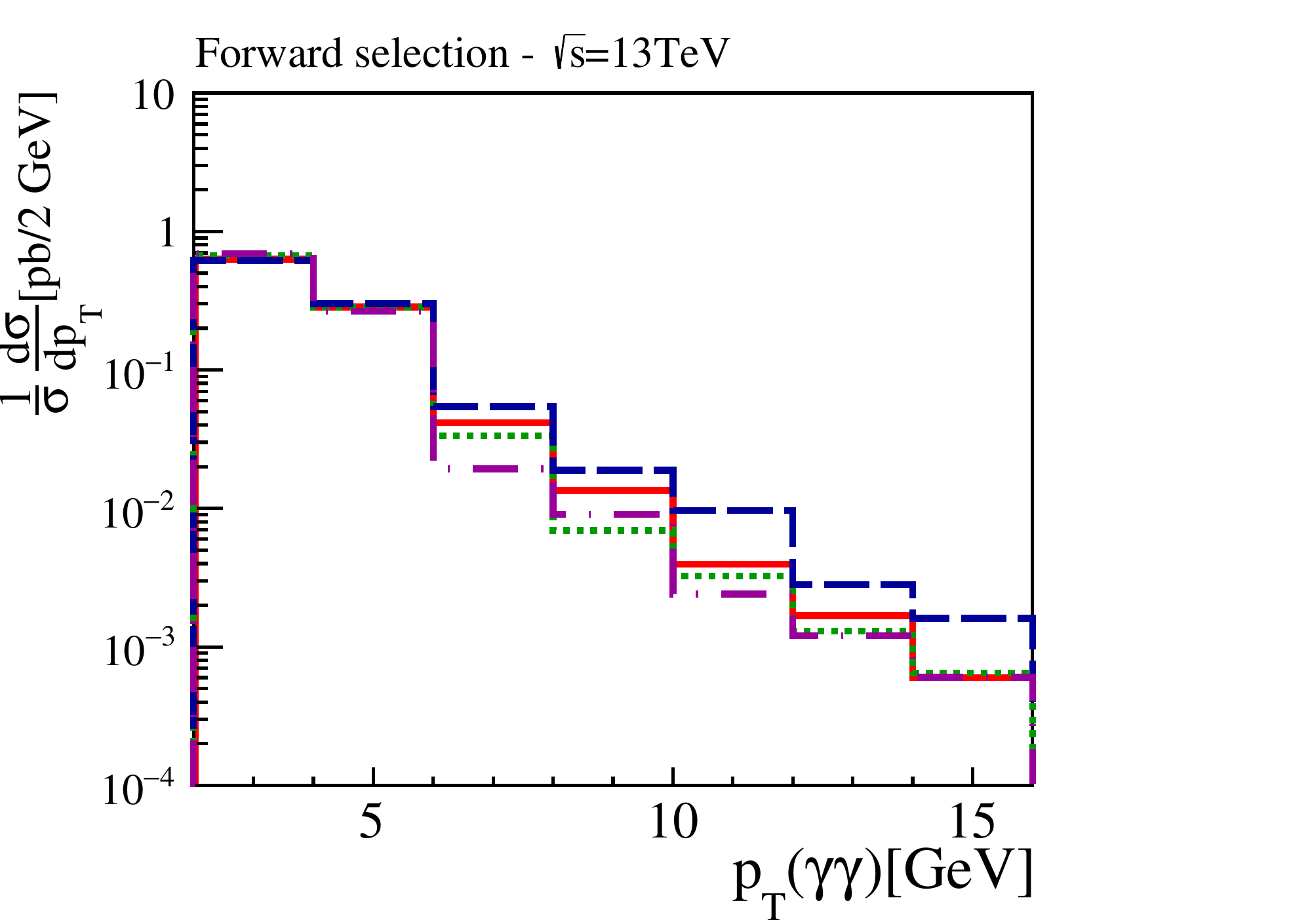}
\includegraphics[width=0.32\textwidth]{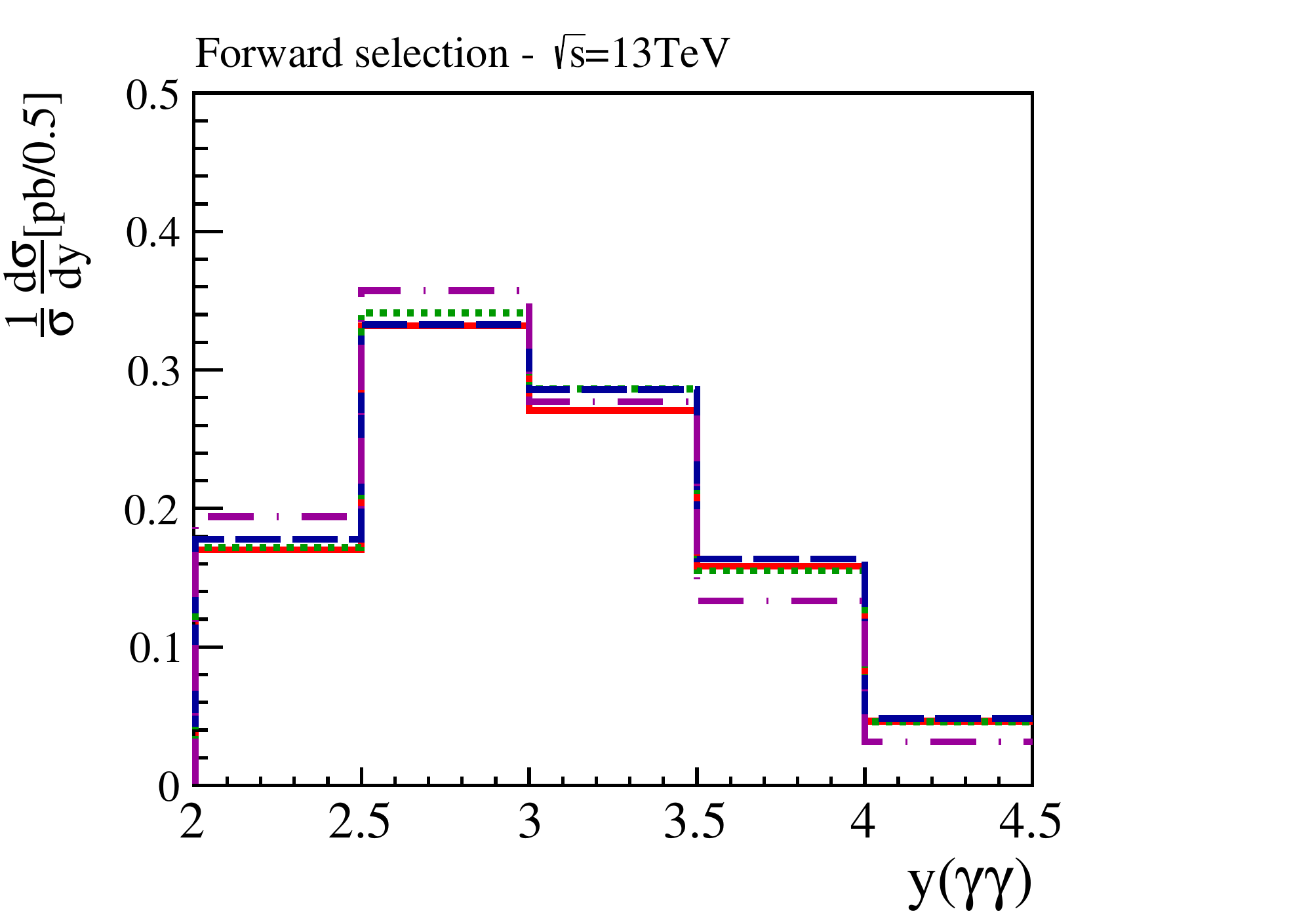}
\caption{Predictions for the normalized invariant mass $m_{\gamma \gamma}$, transverse momentum $p_T({\gamma \gamma})$  and rapidity $y ({\gamma \gamma})$ distributions of the diphoton system produced in $pp$ collisions at the LHC. 
 Results for a central (forward) detector are presented in the upper (lower) panels.
}
\label{fig:central}
\end{figure}
\end{center}

One also has  that the associated predictions are strongly dependent on the diffractive parton distribution considered. 
In particular, the DDP predictions for  central and forward detectors   are a direct probe of the diffractive quark distributions, since  the diphoton production in the kinematical range considered  is dominated by the $q \bar{q} \rightarrow \gamma \gamma$ subprocess, as demonstrated in Fig. \ref{fig:central_gluons}, where we have calculated the distributions using the H1 - Fit A parametrization.   Results for a central (forward) detector are presented in the upper (lower) panels. 
Unfortunately, a direct comparison from our predictions for the total cross sections with future experimental data cannot be used to discriminate between the distinct models for the DPDFs, since these predictions are dependent on the value assumed for $S^2_{{eik}}$. An alternative is to analyze the impact of these different DPDFs on the shape of the differential distributions. 
In Fig. \ref{fig:central} we present our predictions for the  distributions normalized by the associated total cross sections considering the distinct parametrizations and the cuts for  central (upper panels) and forward (lower panels) detectors. An advantage of these normalized distributions is that the predictions are not sensitive to the modeling of the survival factor. One has that the slopes of the invariant mass and transverse momentum distributions  are  sensitive to the modeling of the diffractive quark distribution. Such results indicate that a future experimental analysis of the diphoton production in double diffractive processes can be useful to constrain this distribution.

Two comments are in order. First, one of the main challenges present in the study of diffractive interactions at $pp$ collisions is the experimental separation of these events, especially during the high pile-up running. Pile-up is referred to the multiple soft proton-proton interactions in each bunch crossing of the LHC which  occur simultaneously with the hard process associated to the primary vertex of interest. The average number of pile-up interactions per bunch crossing
during the  Run I to II varies between 20-50 and the expectation for high luminosity LHC is in the range 140-200. As a consequence, the signal events associated to diffractive processes suffer from the background that arise from the other CEP processes with the common final state particles, as those considered in this letter,  and from inclusive processes which are coincided
with the pile-up protons. The presence of forward detectors with high resolution on
momentum and arrival time of protons have been used to suppress background contributions \cite{afp1,afp2,pps}. In principle, the measurement of the forward protons permits to predict the kinematics of centrally produced state, which can be measured separately. The matching between these two measurements can lead to several orders of magnitude suppression in the inclusive background processes.
In addition, the high correlation between the primary vertex
 displacement in the $z$-direction and the arrival time of both tagged protons to the timing forward detectors  present in diffractive and central exclusive processes, can also be used to reduce the inclusive background contribution. Such reduction depends on  the
timing resolution of time of flight detectors. The recent detailed analysis performed in Ref. \cite{nostop}, where a comprehensive study of the $t\bar{t}$ production in exclusive and semi - exclusive processes was presented considering four luminosity scenarios as well as the effect of pile-up background, indicate that there is good prospects for observing the diffractive signal in future measurements. A similar conclusion is expected for the diphoton production. However, a more definitive conclusion deserves a detailed study, as  in Ref.  \cite{nostop}, which we plan to perform  in a forthcoming publication. A second comment is related to the possibility of separation of the diffractive events by the LHCb detector. Such experiment runs at lower instantaneous luminosity, which implies  lower pile-up conditions. Moreover, the presence of the 
HERSCHEL, allow us to suppress  the contribution of the inelastic processes.  Our results for a forward detector demonstrated, the CEP background can be strongly suppressed, and that the diffractive contribution is of the order of pb. Therefore, assuming an integrated luminosity of  3 fb$^-1$, we predict that the number of events per year will be ${\cal{O}}(10^3)$.    Such result indicates that the experimental analysis of the diphoton production by the LHCb is, in principle, feasible.

Finally, let's summarize our main results and conclusions. In this letter we have investigated the diphoton production in diffractive and exclusive  processes present in $pp$ collisions at the LHC. Our main focus was in the possibility of separate the events associated to the double diffractive processes, where the diphotons are produced by the interaction between quarks and gluons of the Pomeron. We have demonstrated that the background associated to the LbL scattering and the exclusive process can be strongly reduced by a cut on the transverse momentum of the diphoton system. As a consequence, the study of the diphoton production with $p_{T}(\gamma\gamma) \ge 3$  GeV becomes a direct probe of the diffractive mechanism and the underlying assumptions associated to the treatment of the gap survival as well as to the description of the Pomeron structure. We shown that the diphoton production is dominated by the $q \bar{q} \rightarrow \gamma \gamma$ subprocess. Moreover, our results indicated that the analysis of the invariant mass, transverse momentum and rapidity distributions are sensitive to the modeling of the diffractive quark distribution.

\begin{center}
\begin{table}[t]
\begin{tabular}{|c|c|c|c|c|c|c|}
\hline 
\hline 
{\bf $pp$ collisions at $\sqrt{s}$ = 13 TeV} & LbL & Durham & DDP(H1-FitA) & DDP(H1-FitB) & DDP(ZEUS) & DDP(H1-ZEUS) \tabularnewline
\hline 
Total Cross section {[}pb{]} & 1.68  & 305.0   & 95.0 & 150.1 & 56.4 & 98.8
\tabularnewline
\hline 
$m_{\gamma\gamma}> 1\:\rm{GeV}, p_{T}(\gamma,\gamma)>0.2\:\rm{GeV}$& 1.28  & 261.7     & 94.8 &149.6& 56.3 &98.2     \tabularnewline
\hline 
$p_{\gamma\gamma}> 3\:\rm{GeV}$&0.0   & 0.0    & 24.9  &  38.2& 12.0 & 18.6     \tabularnewline
\hline
$ 2.0 <\eta(\gamma,\gamma) < 4.5$ &0.0  & 0.0  & 2.8 & 4.2 &1.3 & 1.7
\tabularnewline
\hline 
\hline 
 HERSCHEL  & 0.0 & 0.0  & 1.60   & 2.32  & 0.71 & 0.82  \tabularnewline
\hline
\hline 
\end{tabular}
\caption{Predictions for double diffractive diphoton  cross sections after the inclusion of the exclusivity cuts for a typical forward detector. For comparison the predictions associated to the LbL scattering and exclusive (Durham) process are also presented.}
\label{tab:forward}
\end{table}
\end{center}

\begin{acknowledgments}
VPG acknowledge very useful discussions about diffractive interactions with Marek Tasevsky.
This work was  partially financed by the Brazilian funding
agencies CNPq, CAPES,  FAPERGS, FAPERJ and INCT-FNA (processes number 
464898/2014-5 and 88887.461636/2019-00).
\end{acknowledgments}

\end{document}